\UseRawInputEncoding 
\documentclass[%
 reprint,
 amsmath,amssymb,
 aps,
floatfix,showkeys,floatfix
]{revtex4-1}

\usepackage{graphicx}% Include figure files
\usepackage{dcolumn}% Align table columns on decimal point
\usepackage{bm}% bold math
\usepackage{color}
\usepackage{wrapfig}
\usepackage{physics}
\usepackage{fancyhdr}
\usepackage{lastpage}
\usepackage{amsmath}

\newcommand{\be}{\begin{equation}}
\newcommand{\ee}[1]{\label{#1} \end{equation}}

\def\_#1{\textsubscript{#1}}
\def\^#1{\textsuperscript{#1}}

\usepackage[symbol]{footmisc}

\begin{document}

\preprint{APS/123-QED}

\title{The Effect of Dust and Hotspots on the Thermal Stability of Laser Sails}
\author{Gabriel R. Jaffe$^{\dagger}$ and Gregory R. Holdman$^{\dagger}$}
%\thanks{The authors contribute equally to this paper.}
\affiliation{Department of Physics, University of Wisconsin-Madison, Madison WI 53706 USA}

\author{Min Seok Jang}

\affiliation{
School of Electrical Engineering, Korea Advanced Institute of Science and Technology, Daejeon 34141, Korea
}%

\author{Demeng Feng and Mikhail A. Kats}

\affiliation{
Department of Electrical and Computer Engineering, University of Wisconsin-Madison, Madison WI 53706 USA
}%

\author{Victor Watson Brar}
%\thanks{vbrar@wisc.edu}%
\email[]{vbrar@wisc.edu\newline$^{\dagger}$The authors contribute equally to this paper}
\affiliation{%
Department of Physics, University of Wisconsin-Madison, Madison WI 53706 USA
}%

\date{\today}

\begin{abstract}
\begin{center}
\textbf{\abstractname}
\end{center}
Laser sails propelled by gigawatt-scale ground-based laser arrays have the potential to reach relativistic speeds, traversing the solar system in hours and reaching nearby stars in years. Here, we describe the danger interplanetary dust poses to the survival of a laser sail during its acceleration phase. We show through multi-physics simulations how localized heating from a single optically absorbing dust particle on the sail can initiate a thermal-runaway process that rapidly spreads and destroys the entire sail. We explore potential mitigation strategies, including increasing the in-plane thermal conductivity of the sail to reduce the peak temperature at hotspots and isolating the absorptive regions of the sail which can burn away individually.
\end{abstract}
\maketitle

%\vspace{-10pt}
%\section*{\label{sec:intro}Introduction} 
%\vspace{-10pt}

Space probes propelled by laser sails have been proposed as a means of reaching relativistic speeds and facilitating rapid transit through the solar system and to nearby stars. The Breakthrough Starshot Initiative has laid out one of the most ambitious proposals for such a spacecraft, with the goal of sending a laser sail probe to Proxima Centauri within the next decade.\cite{BreakthroughInitiatives}  In this proposal, a $\sim$10\,m$^2$ sail will be accelerated to a velocity of 0.2$c$ using radiation pressure from a ground-based laser array.  At such a high velocity, the trip from Earth to Proxima Centauri takes only twenty years.  Many recent works have identified materials and optical metasurface structures that can provide the required high reflectivity,\cite{Slovick_Yu_Berding_Krishnamurthy_2013} low areal mass density,\cite{Atwater_Davoyan_Ilic_Jariwala_Sherrott_Went_Whitney_Wong_2018}  and beam-riding capability\cite{Siegel_Wang_Menabde_Kats_Jang_Brar_2019,Ilic_Atwater_2019} required for such a laser sail.  However, thermal management of the sail during its acceleration phase, where the sail will experience laser illumination in excess of 1-10\,GW\,m$^{-2}$ for tens of minutes, remains an outstanding problem.\cite{Atwater_Davoyan_Ilic_Jariwala_Sherrott_Went_Whitney_Wong_2018, ilic_went_atwater_2018,Salary_Mosallaei_2020,Holdman_AOM_2022}

In space, any laser light the sail absorbs can only be dissipated through thermal radiation.  This limitation presents a key design challenge: the sail must be comprised of materials with sufficiently low absorption at the laser wavelength, here chosen to be 1.55\,$\mu$m, but large emissivity at mid- and far-infrared wavelengths (roughly 3--100\,$\mu$m) to enhance radiative cooling.  Designs based on Si nanophotonic structures on SiO$_2$ membranes have been shown to be one attractive geometry, as they can be made highly reflective across the Doppler-broadened laser wavelength bandwidth ($\sim$1.55--1.90\,$\mu$m) while maintaining low mass density and high emissivity, and they can also impart stability on the sail.\cite{Moitra_APL_2014,Siegel_Wang_Menabde_Kats_Jang_Brar_2019,Jin_Li_Orenstein_Fan_2020} We recently showed that Si-based metasurfaces will undergo a thermal runaway process --- with absorption increasing indefinitely with temperature ---- and catastrophically fail if the laser intensity is $>$\,5\,GW\,m$^{-2}$ or the metasurface temperature rises above $\sim$500\,K.\cite{Holdman_AOM_2022}  It is currently not known whether alternative candidate materials to Si, such as silicon nitride,\cite{Steinlechner_PRD_2017,Moura_OE_2018,Ji_APLPho_2021,Lien_OME_2022} also exhibit the temperature- and power-dependent absorption processes of Si that would give rise to a thermal runaway effect, but the optical properties of any material will change at sufficiently high temperatures, as phase changes are inevitable.   The danger posed by thermal runaway necessitates that the sail remain at low temperatures during laser acceleration.  Extrinsic absorption caused by dust or defects in the sail could potentially trigger thermal runaway through localized heating; however, a detailed investigation of this phenomenon has not been previously reported. 

%This thermal runaway process in Si is due two-photon absorption across the Si bandgap, which increases with incident intensity, as well as a dramatic increase in the free carrier absorption at high temperatures.  

% ---------------- Figure 1 ----------------
\begin{figure}[t]
\includegraphics[width = .80\linewidth]{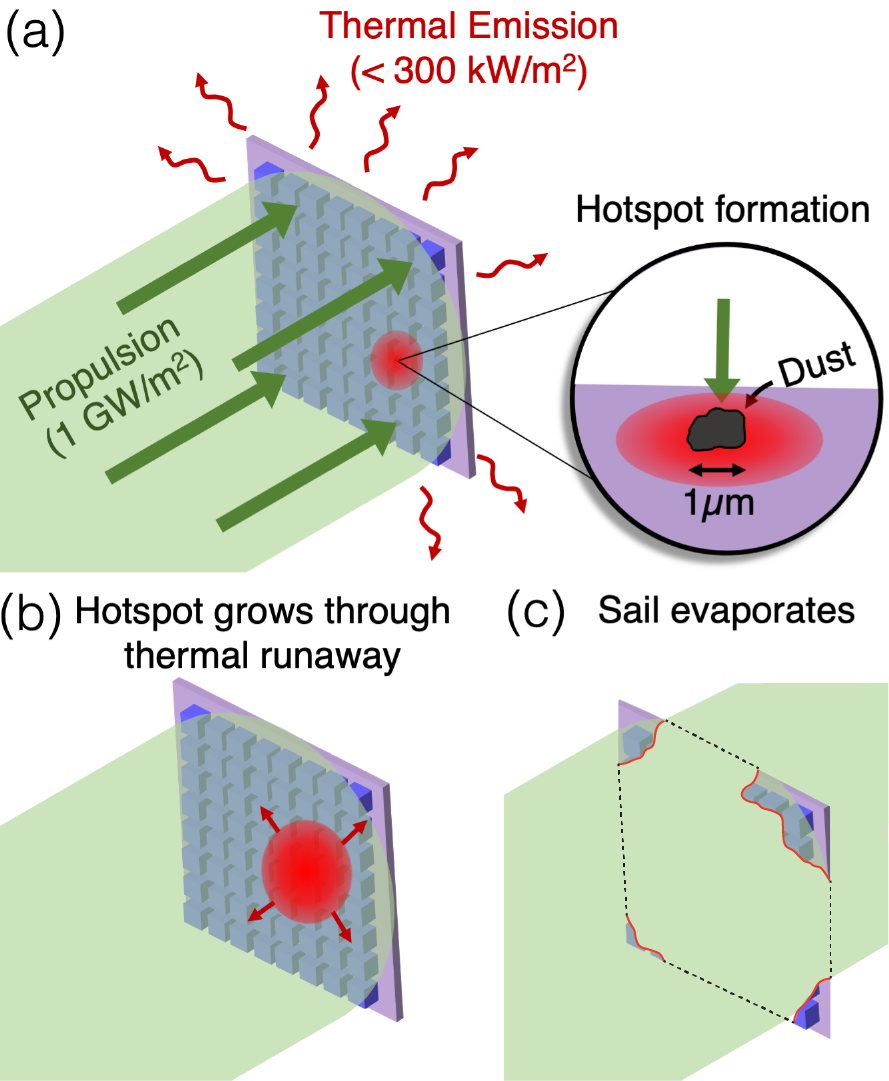}
\caption{a) A 10\,m$^{2}$ laser sail (purple square) made of a Si/SiO$_2$ metasurface is accelerated through space by the radiation pressure from a 10\,GW 1.55\,$\mu$m laser (green arrows).  A 1\,$\mu$m zodiacal dust particle on the sail absorbs the impinging laser light and creates a hotspot on the sail. b) The hot area of the sail grows through a thermal runaway process wherein the Si absorbs more laser energy than can be locally thermally emitted by the sail. c) The sail is ultimately vaporized.
}
\label{fig:2}
\end{figure}
% -------------------------------------------

 In this article, we quantify the danger that dust and subsequent hotspots pose to laser sails under high-power illumination.  Our simulations show that conductive heat spreading from the local absorption of a single micron-sized dust particle can trigger a thermal runaway process that ultimately destroys the entire sail. These results highlight the important role that the nanoscale thermal-transport characteristics of sail structures play in determining the sail's thermal stability during laser acceleration.  To aid in future sail designs, we propose two strategies for reducing the risk of thermal runaway due to dust: (1) increasing the in-plane thermal conductance of the sail to reduce the hotspot temperatures, and (2) dividing the sail into isolated regions that can burn away individually to prevent thermal runaway from consuming the rest of the sail. The mitigation strategies outlined here are generally applicable to all radiatively cooled high-power optics that exhibit a thermal runaway behavior at high temperatures.

The inevitable interaction between laser sails and absorptive particles will occur because the earth is surrounded by the zodiacal dust cloud.\cite{Lasue_PSS_2020}  At a solar radius of 1\,AU, the zodiacal dust is expected to be $\sim$25-50\% carbonaceous compounds by mass and have a space number density density of 6.34$\times$10$^{-6}$\,m$^{-3}$ for particles ranging in radius from 10 to 1000\,nm.\cite{Hadamcik_PSS_2020,Grun_Book_2001}  A 10\,m$^2$ sail in a similar orbit to Earth will experience an impact with one of these dust particles every $\sim$3\,s with impact velocities assumed to be near 20 km\,s$^{-1}$.\cite{Grun_ICA_1985} Damage assessments of hypervelocity meteoroid impacts on thin metal plates suggest that much of the sub-micron dust at this impact velocity will embed into the sail rather than passing straight through.\cite{NASA_MeteorDamage} 

Dust particles present on the sail surface or embedded in the sail will form small hotspots as they absorb energy from the laser and spread that energy into the sail through heat conduction (Fig.\ref{fig:2}a). Even in an optimistic case, assuming a sail with an extremely high reflection coefficient of 99.9\% and an incident beam of 1\,GW\,m$^{-2}$, the transmitted laser flux through the sail will be equivalent to the total thermal radiated power for a blackbody at a temperature $>$\,2000\,K calculated using the Stefan-Boltzmann Law. Thus, no matter where on the sail the dust is located (front or back), it has the potential to form hotspots well above the thermal-runaway temperature of the metasurface.  Hotspots could also be formed through high velocity impacts with dust during the sail's acceleration phase.  A sail of area $A$, accelerated across a distance of $D$ through a dust cloud with particle density $\rho$ will encounter $N = A\cdot D\cdot\rho$ particles.  A 10\,m$^2$ sail will therefore experience impacts with $\sim$10$^6$ particles while being accelerated across a distance of 0.1\,AU assuming the dust density listed above. 

%\vspace{-10pt}
%\section*{\label{sec:results}Results and Discussion} 
%\vspace{-10pt}

\begin{figure*}
\includegraphics[width = .6\linewidth]{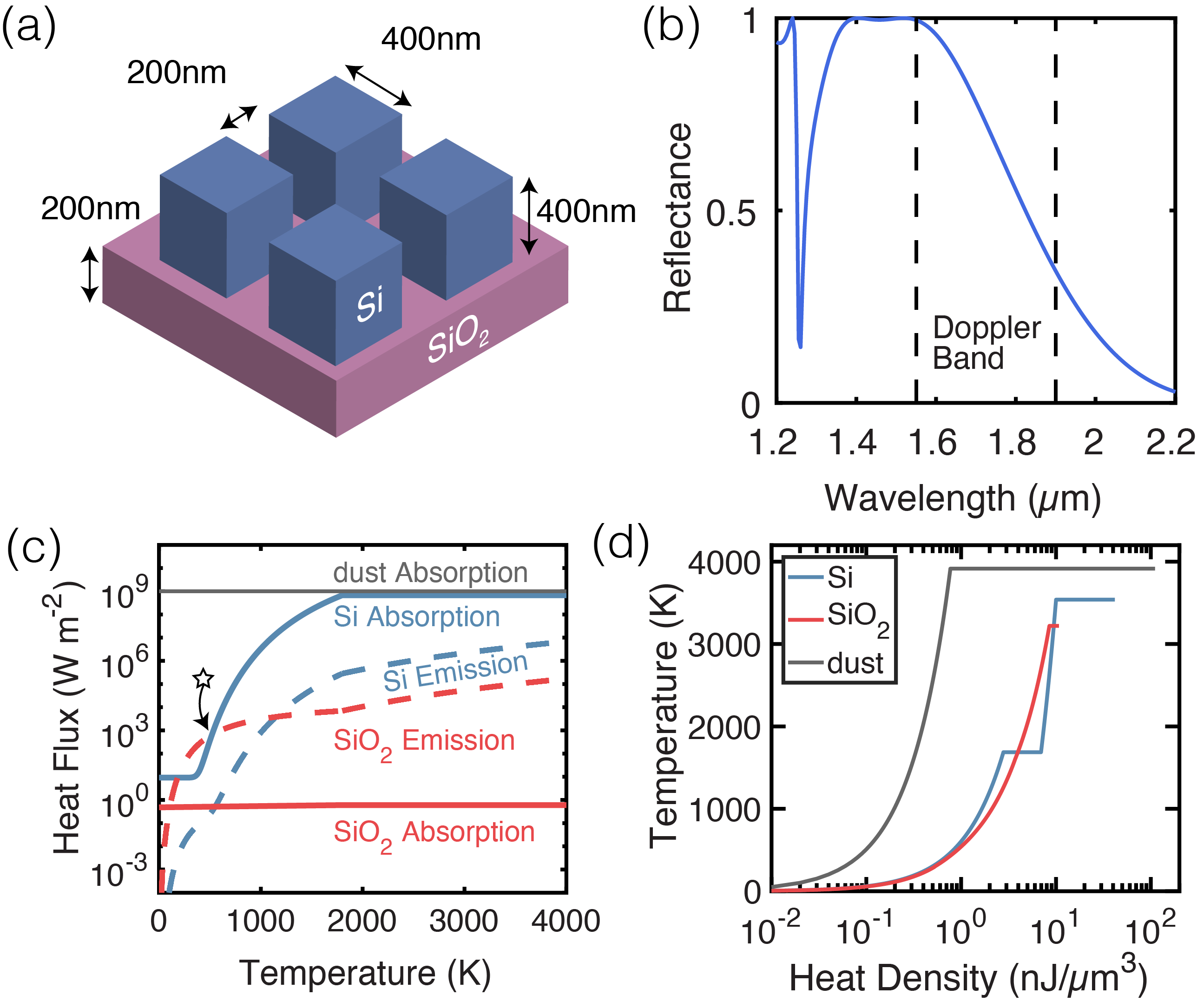}
\caption{(a) A diagram of two periods of a metasurface laser sail used for the thermal simulations reported here.  It is composed of 400\,nm Si blocks with a pitch of 600\,nm on a 200\,nm thick SiO$_2$ substrate. Note, the dimensions of the metasurface were altered slightly from those used for the reflectance and optical absorption calculations reported in Ref.\,\cite{Holdman_AOM_2022} in order fit evenly on a 200\,nm pitch simulation grid for the heat-transport simulations. (b) The reflectance of the metasurface is shown in blue. The doppler-broadened laser bandwidth is denoted by dashed lines. (c) The absorbed laser power and thermally radiated power per unit area of the Si (blue) and SiO$_2$ (red) regions of the metasurface as a function of temperature assuming a normally incident beam of 1\,GW\,m$^{-2}$ of 1.55\,$\mu$m light.  In this work, the zodiacal dust is assumed to be made entirely of amorphous carbon and act as a perfect blackbody absorber (grey line). At low temperatures, the SiO$_2$ (red dashed line) provides the majority of the emission but it is overtaken by Si (blue dashed line) above $\sim$1150\,K.  At all temperatures, silicon (solid blue line) dominates the absorption.  Above the thermal runaway temperature of 516\,K (denoted by a black star), the sail absorbs more energy than it can emit, and undergoes thermal runaway. (d) Literature models of the relationship between temperature and volumetric heat capacity of SiO$_2$ (red), Si (blue), and amorphous carbon (grey).\cite{Hoch_JPC_1955,Richet_GCA_1982}
}
\label{fig:1}
\end{figure*}

We study the effect an optically absorbing dust particle would have on the high-reflectivity Si/SiO$_2$ metasurface laser sail design previously published in Ref.\,\cite{Holdman_AOM_2022} and depicted here in Fig.\,\ref{fig:1}a.  This structure acts as a near perfect reflector across the Doppler-broadened laser bandwidth (Fig.\,\ref{fig:1}b). While this structure does not provide any beam-riding stability to the laser sail, the conclusions of this work should generalize to the more complex beam-riding metasurface structures,\cite{Siegel_Wang_Menabde_Kats_Jang_Brar_2019} and those optimized for acceleration efficiency.\cite{Jin_Li_Orenstein_Fan_2020} Calculations of the temperature-dependent absorption and emission of the Si pillars and SiO$_2$ substrate assuming a 1\,GW\,m$^{-2}$ incident laser flux at 1.55\,$\mu$m are shown in Fig.\,\ref{fig:1}c.  Detailed descriptions of these calculations can be found in Ref.\,\cite{Holdman_AOM_2022}.  The dimensions of the metasurface were altered slightly from Ref.\,\cite{Holdman_AOM_2022} in order fit evenly on a 200\,nm pitch simulation grid for the heat-transport simulations. The thermal runaway of this metasurface, where the absorption becomes larger than the emission, begins at a temperature of $\sim$500\,K and can be identified in Fig.\,\ref{fig:1}c where the Si absorption (solid blue line) crosses SiO$_2$ emission (dashed red line). Literature models of the temperature-dependent heat capacity, including the heat of condensation and vaporization for each material are shown in Fig.\,\ref{fig:1}d.  The dust, a large fraction of which is composed of carbon, is assumed to be a perfect blackbody absorber with negligible emission and have the heat capacity of amorphous carbon.  These simplifications are justified because: 1) the dust does not have significant surface area to radiate much heat relative to the underlying metasurface and 2) the incident laser intensity is so immense (the equivalent radiated power of a blackbody at $\sim$12,000\,K) that changing the absorptivity of the dust would only slightly alter the time it takes it to reach its sublimation temperature.

 % ---------------- Figure 2 ----------------
% -------------------------------------------

\begin{figure*}
\includegraphics[width = 1.0\linewidth]{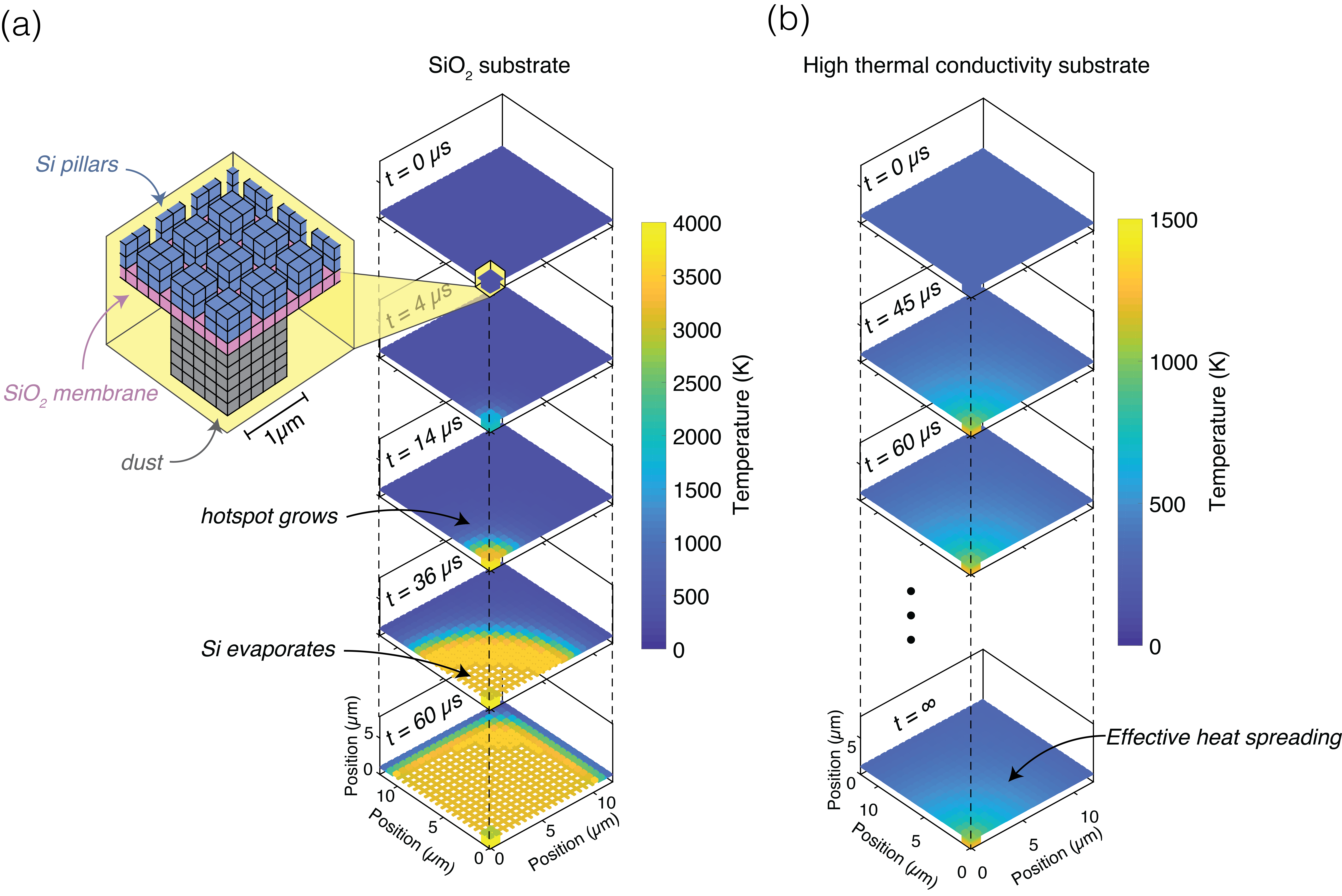}
\caption{Finite-difference time-domain heat transport simulations of one quadrant of the metasurface depicted in Fig.\,\ref{fig:1}, consisting of Si pillars on a membrane of SiO$_2$ initially at 300\,K with 1 GW\,m$^{-2}$ incident laser at 1.55\,$\mu$m.  One quadrant of a 2\,$\mu$m wide piece of amorphous carbon dust is present underneath the SiO$_2$ supporting layer centered at the origin. The dust is assumed to be a perfect blackbody absorber with the heat capacity of amorphous carbon (Fig.\,\ref{fig:1}d) and the absorption and emission of the Si/SiO$_2$ metasurface follows the temperature-dependent calculations shown in Fig\,\ref{fig:1}c. The outer boundaries of the metasurface in the positive x and y directions are kept at a constant temperature of 300\,K. a) The temperature of the metasurface and dust at successive time snapshots. Initially, the dust particle absorbs energy from the incident laser beam which spreads through heat conduction to the rest of the metasurface.  As the temperature of the Si pillars around the dust cross the thermal runaway temperature ($t = $ 14\,$\mu$s), the total energy being absorbed into the sail increases, accelerating the thermal runaway process.  Before the dust or Si is able to evaporate, new nearby regions of Si are heated above the thermal runaway temperature, causing the hotspot to grow without bound towards the edge of the metasurface.  b)  A simulation where the heat-spreading capability of the metasurface has been improved by increasing the in-plane thermal conductivity used for the SiO$_2$ layer by a factor of 8 to 11.2 W\,m$^{-1}$\,K$^{-1}$.  In this case, the in-plane heat conduction is sufficient to quickly spread out the energy absorbed by the dust particle and the neighboring Si pillars do not enter thermal runaway.
}
\label{fig:3}
\end{figure*}

Figure\,\ref{fig:3}a shows a simulation of dust initiating thermal runaway in an Si/SiO$_2$ metasurface sail. The modeled geometry includes one 12$\times$12\,$\mu$m quadrant of an Si/SiO$_2$ metasurface with one quadrant of a 2\,$\mu$m wide 1\,$\mu$m tall dust particle located at the origin against the bottom of the SiO$_2$ layer. An analysis of how dust particle size affects our findings is provided in the supplement. Only one quadrant of the metasurface and dust particle are simulated because the structure is symmetric. Even if the sail were initially cooled before launch, our previous work showed that Si/SiO$_2$ metasurfaces will rapidly heat to an equilibrium temperature between 200-400\,K, depending on material quality.\cite{Holdman_AOM_2022} We therefore begin the simulation with the sail and the dust at 300\,K.  The outer boundary of the metasurface is kept fixed at 300\,K to approximate the effects of heat conduction and subsequent thermal radiation from areas of the sail outside the simulation bounds. We use finite difference time-domain (FDTD) simulations to simultaneously model the optical absorption and local temperature of the sail, with a characteristic grid dimension of 200nm, and a time resolution of 100ps (see Methods). Initially, at $t = $ 0\,$\mu$s, only the dust absorbs energy from the laser.  After 14\,$\mu$s, the energy absorbed by the dust particle has raised the temperature of the neighboring Si pillars to the point of thermal runaway. The Si pillars begin to absorb additional energy and conduct heat outwards, spreading the hotspot. As they continue to heat, the Si  melts and the supporting SiO$_2$ gradually softens. Eventually, the SiO$_2$ directly beneath each pillar evaporates because the evaporation temperature of SiO$_2$ is slightly below that of Si. After the supporting SiO$_2$ has evaporated, the molten Si pillars are left thermally isolated from the rest of the sail, and, without a heat conduction pathway to dissipate heat, evaporate shortly thereafter. The result is a mesh of semi-liquid SiO$_2$ as seen starting at 36\,$\mu$s in Fig.\,\ref{fig:3}a. The hotspot will continue to grow indefinitely until it reaches the edge of the sail because the Si pillars at the edge of the hotspot absorb sufficient energy into the sail to increase the hotspot size before evaporating. A necessary simplification made here is that the metasurface geometry is made of rigid bodies with no fluid motion, however, the Si and SiO$_2$ undergo liquid phase transitions before evaporating.  While beyond the scope of this work, a detailed investigation of the behavior of these metasurfaces when partially molten and under large optical forces would be of great interest.

% ---------------- Figure 3 ----------------
% -------------------------------------------

% ---------------- Figure 4 ----------------
\begin{figure*}
\includegraphics[width = 1.0\linewidth]{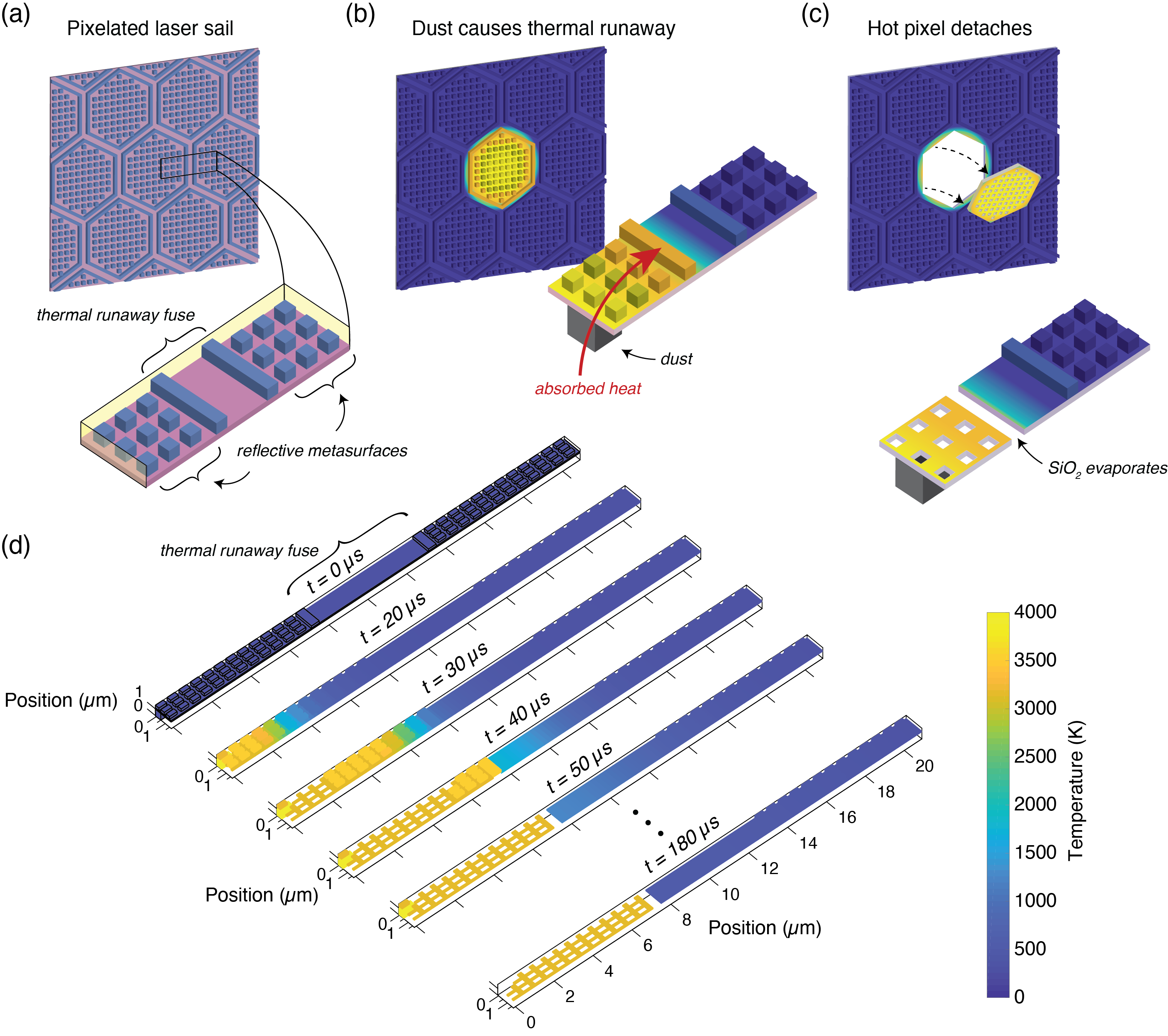}
\caption{(a) A schematic diagram of a laser sail composed of an array of small reflective segments.  Each segment consists of an array of Si resonators (blue cubes) on a SiO$_2$ substrate (pink).  Each segment is surrounded by a continuous strip of Si that serves as a ``thermal runaway fuse''.  (b)  A piece of dust initiates thermal runaway in a segment, which begins to rapidly heat and evaporate.  (c) The Si melts the supporting SiO$_2$ in a ring around the hot segment which then detaches and is thermally isolated from the surrounding metasurface. (d) Time snapshots from a finite difference time domain simulation of a 600\,nm wide and 21\,$\mu$m long strip spanning the gap between two reflective elements in the segmented metasurface shown in (a).  A 1\,$\mu$m wide piece of dust is present underneath the SiO$_2$ at the origin.  The metasurface begins the simulation at 300\,K and the end of the strip opposite the dust is held at a constant temperature.
}
\label{fig:4}
\end{figure*}
% -------------------------------------------

A promising mitigation strategy to prevent hotspot-induced thermal runaway is to improve the heat-spreading ability of the metasurface by increasing its in-plane thermal conductance.  If the heat is spread out quickly, the region of the sail surrounding the dust will remain below the thermal runaway temperature and the energy absorbed by the dust can be radiated away safely over a large area.  This effect is simulated in Fig.\,\ref{fig:3}b, where the supporting SiO$_2$ layer has been replaced by a theoretical material with identical optical properties but with a thermal conductivity of 11.2\,W\,m$^{-1}$\,K$^{-1}$, which is a factor of 8 above its intrinsic value.  The in-plane thermal conductance of the metasurface is now large enough to conduct the absorbed heat near the dust away to the simulation boundaries without allowing the hotspot size to increase.  It is important to note that the constant temperature boundary condition imposed here likely overestimates the thermal conductance between the simulated metasurface area and the relatively large thermal mass of the rest of the sail.  As such, the increased in-plane conductivity of the supporting layer used here should be viewed as an optimistic lower bound of the conductivity needed to achieve sufficient heat spreading for this metasurface.

We note that there are many materials with large in-plane thermal conductivities that could be incorporated into reflective metasurface designs to improve heat spreading.  Near room temperature, graphene, diamond, boron arsenide, and hexagonal boron nitride have thermal conductivities of 5300, 2200, 1300, and 400\,W\,m$^{-1}$\,K$^{-1}$, respectively, and would provide 100-1000 times more in-plane thermal conductance than SiO$_2$.\cite{Fu_2DMat_2019,Slack_JAP_1964,Kang_Science_2018,Sichel_PRB_1976}  The thermal conductivity of SiN$_x$ membranes has been found to be as high as $\sim$10\,W\,m$^{-1}$\,K$^{-1}$, which still represents a factor of 7 improvement over SiO$_2$.\cite{Sikora_RSI_2012}  The thermal conductivity of Si is also a factor of $\sim$100 larger than SiO$_2$, meaning that a reflective metasurface design that included a continuous film of Si, rather than the thermally isolated pillars in the design considered in this work, could show significantly higher in-plane thermal conductance. As an example, the addition of a 6nm thick hexagonal boron nitride film would be sufficient to achieve the needed in-plane thermal conductance and would only increase the areal mass density of the metasurface by $\sim$1.3\%.

A second mitigation strategy for preventing dust-driven thermal runaway would be to segment the metasurface sail into individual pixels that could burn away individually, thereby preventing the hotspot from spreading to neighboring segments.  In this scheme, pictured in Fig.\,\ref{fig:4}a, each pixel would contain a reflective metasurface surrounded by a ``thermal runaway fuse''.  When the ring of Si surrounding each pixels heats up above 3220\,K, the supporting SiO$_2$ will evaporate and the pixel will become detached from the metasurface, as is pictured in Fig.\,\ref{fig:4}c.  A time-resolved simulation of the thermal runaway fuse is shown Fig.\,\ref{fig:4}d, where a three-period-wide strip of metasurface that spans the gap between two reflective metasurfaces is simulated.  A 1\,$\mu$m wide piece of dust is placed on the bottom of the SiO$_2$ layer on the left side of the strip while the opposite edge of the strip is kept at 300\,K throughout the simulation to approximate heat flowing to regions outside the simulation bounds.  From $t = $ 0 to 40\,$\mu$s, the dust causes a thermal runaway that rapidly spreads through the metasurface.  At $t = $ 50\,$\mu$s, the hot segment has become thermally isolated because the supporting SiO$_2$ has been evaporated across the width of the strip.  Within the next 130\,$\mu$s, the small amount of heat that traveled to the neighboring pixel has radiated away and the pixel temperature approaches 300\,K.

The larger the separation between segments where no Si is present, the less heat will be able to escape from the hot pixel before it becomes detached.  The trade-off with this strategy is that the larger the gap size, the smaller the fraction of the sail that can be made reflective.  It is also important to consider the number of pixels in the sail versus the expected number of dust impacts.  In order for 90\% of the segments to survive the acceleration phase after the 10$^6$ dust impacts predicted for the sail dimensions stated above, the sail would need to contain 10$^7$ segments, each with an area of 1\,mm$^2$.  This assumes the most pessimistic case where each dust impact destroys a segment.  The areal fraction of fuse region to reflective metasurface in this scheme would be 1\,\%.

This thermal isolation strategy could also be implemented by thinning the substrate to decrease its in-plane thermal conductance or using a different substrate material that has a smaller thermal conductivity than SiO$_2$.  The thermal conductance between two Si pillars across the thermal runaway fuse shown in Fig.\ref{fig:4} is $\sim$35\,nW\,K$^{-1}$.  Reducing the SiO$_2$ layer thickness to 8\,nm would be sufficient to achieve this low thermal conductance.  For a 200\,nm thick substrate, materials with thermal conductivity of $<$0.058\,W\,m$^{-1}$\,K$^{-1}$ would need to be used.  While traditional semiconductors and oxides are limited to conductivities $>$1\,W\,m$^{-1}$\,K$^{-1}$, polymers have conductivities $\sim$0.1\,W\,m$^{-1}$\,K$^{-1}$, and dense fullerene-based organic compounds and organic-ligand-stabilized inorganic nanocrystal arrays have been synthesized with conductivities of 0.03-0.06\,W\,m$^{-1}$\,K$^{-1}$.\cite{Qian_NM_2021}

SiN is likely the leading candidate material for laser sails beyond silicon and silicon oxide. Unfortunately, there has not been a comprehensive study measuring the temperature and power-dependent absorption coefficients of SiN, which makes it unfeasible to calculate thermal runaway effects and subsequent sail evaporation.  At elevated temperatures, SiN dissociates into Si and N$_2$, which suggests that if a SiN nanomembrane were made hot enough, a similar thermal runaway behavior could occur.\cite{Singhal_CI_1976}   For example, if such a dissociation process were to create a small semi-molten Si region near a dust particle or defect, the high absorption of the Si would heat up the surrounding SiN, creating an even larger area of Si and in this manner the spread of thermal runaway could become self-sustaining.  The mitigation strategies highlighted here; high in-plane thermal conductivity substrates to reduce hotspot peak temperatures, using thermal runaway fuses to isolate section of the metasurface, would also be applicable in the case of a SiN membrane.

%\vspace{-10pt}
%\section*{\label{sec:discussion}Conclusions} 
%\vspace{-10pt}

The zodiacal dust cloud presents a direct danger to the thermal stability of laser sails under illumination.  In particular, absorptive dust particles can embed themselves in the sail before launch or heat regions of the sail through high velocity impacts during laser acceleration.  The resulting hotspots have the potential to grow and cause the entire sail to evaporate, as demonstrated here for a reflective Si/SiO$_2$ metasurface.  Designing sails to have a large in-plane thermal conductance can reduce hotspot temperatures and prevent thermal runaway by spreading the absorbed heat from the dust particle over larger areas.  Additionally, we demonstrate how segmenting the sail into small reflective elements that can burn away individually can prevent the spread of thermal runway.  These results pave the way for laser sail designs that are thermally stable during acceleration in the dusty vacuum of space.

%\vspace{-10pt}
%\section*{\label{sec:results}Methods} 
%\vspace{-10pt}

We model the time evolution of the temperature of a Si/SiO2 laser sail illuminated with a 1.55\,$\mu$m laser with an incident power density of 1\,GW\,m$^{-2}$ using 3D finite-difference time-domain (FDTD) heat transport simulations.  The simulation grid consists of 200\,nm cubic cells that represent small regions of the metasurface.  During each 100\,ps simulation time step, the thermally radiated power and absorbed laser power is calculated for each grid point following the temperature-dependent calculations shown in Fig.\,\ref{fig:1}c. Above 1800\,K, we assume the absorption of Si and SiO$_2$ are constant with temperature and that the emission of each material follows the Stefan-Boltzmann Law with an emissivity of 0.473 for Si and 0.0117 for SiO$_2$.  This approximation is necessary because our models rely on extrapolations from literature models that were verified at temperatures $<$\,1200\,K. Above 1800\,K we find that these extrapolations yield unphysical values, such as the absorptivity of Si exceeding 1. Our predicted Si emitted power density agrees well with experimental data at its melting temperature of 1687\,K, which to our knowledge is the highest temperature data available.\cite{Rhim_IJT_1997}  To approximate volumetric heating, the areal heat flux shown in Fig.\,\ref{fig:1}c is evenly distributed along the grid points in the cross-plane direction for each material.  The temperature of each grid point is calculated from its heat density according to the material models in Fig.\,\ref{fig:1}d.  At each time step, heat is allowed to flow between nearest neighbor grid points, taking into account each material's individual thermal conductivity.  When a grid point exceeds the evaporation temperature of the material it represents, the point is removed from the simulation. The SiO$_2$ is assumed to have a thermal conductivity of 1.4\,W\,m$^{-1}$\,K$^{-1}$, Si 140\,W\,m$^{-1}$\,K$^{-1}$, and the dust 2\,W\,m$^{-1}$\,K$^{-1}$. The heat of fusion, vaporization for Si and SiO$_2$ as well as the heat of sublimation of amorphous carbon are included in the temperature-dependent heat capacity models for the materials as shown in Fig.\,\ref{fig:1}d.  Note that SiO$_2$ undergoes a gradual softening to a liquid phase and does not have a distinct heat of fusion. 

\vspace{-10pt}
\section*{\label{sec:results}Acknowledgments} 
\vspace{-10pt}

This work is supported by the Breakthrough Initiatives, a division of the Breakthrough Prize Foundation, and by the Gordon and Betty Moore Foundation through a Moore Inventors Fellowship. This work was also funded, in part, by NSF grant \#1750341.

\bibliographystyle{achemso_custom}
\bibliography{Bibliography}
\end{document}

\title{Supporting Information\\
\small The Effect of Dust and Hotspots on the Thermal Stability of Laser Sails}

\author{Gabriel R. Jaffe$^{\dagger}$}
%\thanks{The authors contribute equally to this paper.}
\affiliation{Department of Physics, University of Wisconsin-Madison, Madison WI 53706 USA}

\author{Gregory R. Holdman$^{\dagger}$}
%\thanks{The authors contribute equally to this paper.}
\affiliation{Department of Physics, University of Wisconsin-Madison, Madison WI 53706 USA}

\author{Min Seok Jang}
\affiliation{
School of Electrical Engineering, Korea Advanced Institute of Science and Technology, Daejeon 34141, Korea
}%

\author{Demeng Feng}
\affiliation{
Department of Electrical and Computer Engineering, University of Wisconsin-Madison, Madison WI 53706 USA
}%

\author{Mikhail A. Kats}
\affiliation{
Department of Electrical and Computer Engineering, University of Wisconsin-Madison, Madison WI 53706 USA
}%

\author{Victor Watson Brar*$^,$}
%\thanks{vbrar@wisc.edu}%
%\email[]{}
\affiliation{%
Department of Physics, University of Wisconsin-Madison, Madison WI 53706 USA
}%

\date{\today}

\maketitle

% this adds page number to title page
\thispagestyle{fancy}

% centers and displays the email line under the affiliations
\onecolumngrid
\begin{center}
*Email: vbrar@wisc.edu\newline$^{\dagger}$The authors contribute equally to this paper\clearpage
\end{center}

%=====================================
\section{The effect of dust particle size}

The size and areal density of dust particles is an important factor in determining whether thermal runaway occurs.  Following a scaling argument described below, we estimate that isolated dust particles with radii $<$350\,nm will not initiate thermal runaway.  This limit is supported by simulations using different dust sizes, plots of which can be seen below.  Furthermore, we calculate analytically that the total areal coverage of dust of a 10\,m$^2$ sail must stay below 1\,mm$^2$ to prevent thermal runaway.

In the first simulation shown below, the dust particle radius is 200\,nm and no thermal runaway is observed.  In the second simulation, the dust particle has a radius of 400\,nm and thermal runaway destroys the metasurface.  

% ---------------- Figure S1 ----------------
\begin{figure}[h]
	\centering
	\let\nobreakspace\relax
	\renewcommand{\figurename}{Figure S}
\includegraphics[width = 1\linewidth]{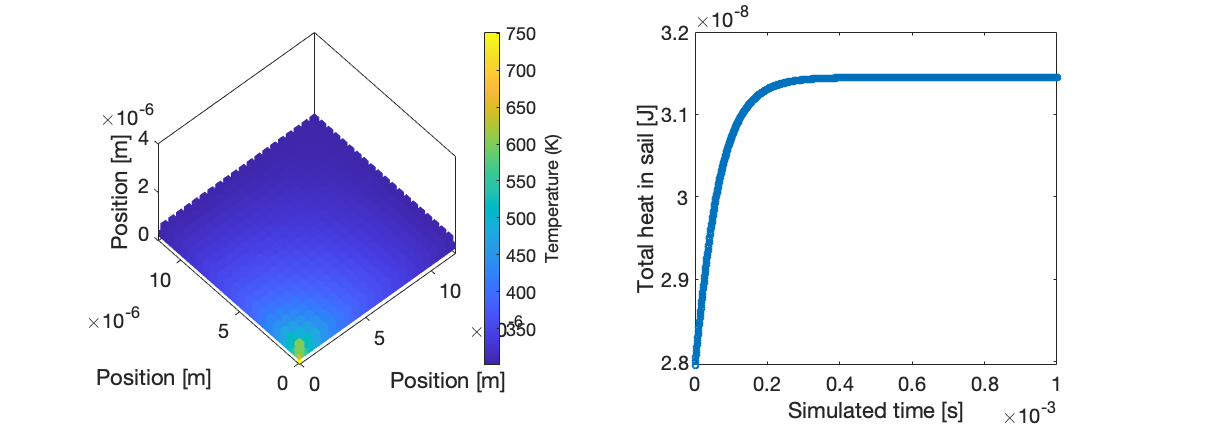}
\caption{Simulation of the same metasurface depicted in Fig. 3a of the main text but with a dust particle with a radius of 200nm.  The particle is small enough that thermal runaway does not spread.  The left plot shows a 3D temperature map of the metasurface after 1ms.  The right plot shows the total heat in the simulated volume as a function of time which appears to have reached a steady state.
}
\label{fig:S1}
\end{figure}
% -------------------------------------------

% ---------------- Figure S2 ----------------
\begin{figure}[h]
	\centering
	\let\nobreakspace\relax
	\renewcommand{\figurename}{Figure S}
\includegraphics[width = 1\linewidth]{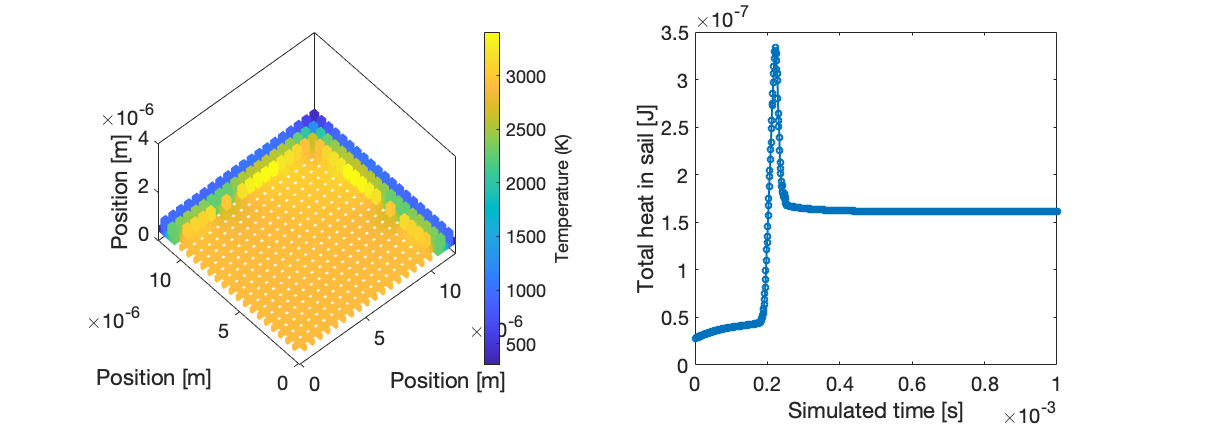}
\caption{Simulation of the same metasurface depicted in Fig. 3a of the main text but with a dust particle with a radius of 400nm.  The particle is large enough that thermal runaway destroys the metasurface, leaving predominantly a mesh of molten SiO2.  The left plot shows a 3D temperature map of the metasurface after 1ms.  The right plot shows the total heat in the simulated volume as a function of time.  The onset of thermal runaway can be seen around 200µs, where the rate of heat absorption dramatically increases.  The total heat in then begins to decrease as metasurface material evaporates.
}
\label{fig:S2}
\end{figure}
% -------------------------------------------

% ---------------- Figure S3 ----------------
\begin{figure}[h]
	\centering
	\let\nobreakspace\relax
	\renewcommand{\figurename}{Figure S}
\includegraphics[width = .50\linewidth]{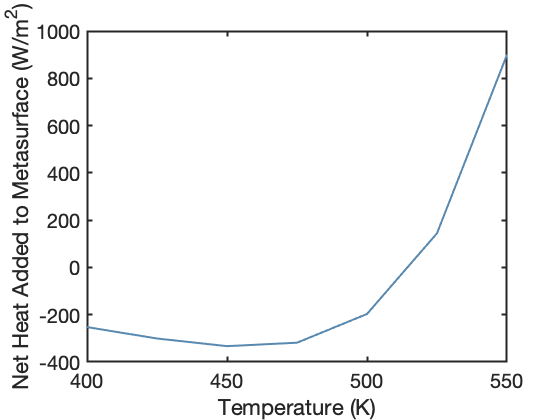}
\caption{The total absorbed heat minus the total emitted heat as a function of temperature of the metasurface calculated from Fig. 2c of the manuscript.  Positive fluxes represent net absorption and negative values represent net emission of the metasurface.
}
\label{fig:S3}
\end{figure}
% -------------------------------------------

Under steady state conditions, the temperature rise of a Si pillar with a fixed heat input from a dust particle can be characterized as $T_{pillar}= R \cdot P$, where $P = \pi a^2 \cdot 1\,GW m^{-2}$   is the heat absorbed by the dust particle of radius $a$, and $R$ is thermal resistance, units of K W$^{-1}$, between a single Si pillar and the surrounding metasurface. In Fig.\,3b of the main text we see that increasing the in-plane thermal conductivity of the substrate by a factor of 8, and as a result, decreasing $R$ by the same factor, is sufficient to keep $T_{pillar}$ small enough to prevent thermal runaway from spreading.  Reducing the particle radius by a factor of $\sqrt{8}$ should have the same effect of as decreasing $R$ by a factor of 8.  Using this approach, we estimate that isolated particles with radii $<$\,350\,nm will not cause thermal runaway for the case of a 200\,nm thick SiO$_2$ substrate.

If we assume a homogenous areal distribution of dust across the entire sail, we can analytically calculate the minimum areal coverage of dust that would cause thermal runaway. Plotted below, we show the heat flux absorbed minus the heat emitted for the metasurface calculated from the curves shown in Fig. 2c of the manuscript.  We see that the maximum net radiated heat flux of the metasurface is $\sim$300\,W\,m$^{-2}$ which occurs at $\sim$\,450\,K.  If the absorbed power flux of the dust particles on the metasurface is greater than this value, the sail will be unable to sufficiently cool radiatively and will always enter thermal runaway.  Dust areal coverage of 1\,mm$^{2}$ on a 10\,m$^2$ sail would absorb the necessary 3000\,W to push the sail into thermal runaway.  Alternatively, this corresponds to a 7.5\,nm particle on one out of every 1600 Si pillars.